\documentclass{amsart}

\usepackage{amsmath,amsxtra,amsthm,amssymb,amsfonts,graphicx,subcaption} 

\usepackage[pdfusetitle,bookmarks=true,bookmarksnumbered=false,bookmarksopen=false,breaklinks=false,pdfborder={0 0 0},backref=false,colorlinks=true]{hyperref}

\def\l{\left}
\def\r{\right}
\def\as{\!\mathrel{\mathop:}=}

\def\nat{\mathbb{N}}
\def\rls{\mathbb{R}}
\newcommand{\rec}[1]{\frac{1}{#1}}
\newcommand{\mat}[2]{\rls^{#1 \times #2}}

\def\lib{\texttt{ExB SVD Library}}
\def\libs{\texttt{ExB SVD Library}\;}

\begin{document}
\title[Out-of-core SVD]{Out-of-core singular value decomposition}

\author[V. Demchik]{Vadim Demchik}
\author[M. Ba\v{c}\'ak]{Miroslav Ba\v{c}\'ak}
\author[S. Bordag]{Stefan Bordag}
\date{\today}

\address{ExB Labs, Seeburgstr. 100, 04103 Leipzig, Germany}
\email{demchik@exb.de}
\email{bordag@exb.de}

\date{\today}
\subjclass[2010]{Primary: 65F30, 65Y05. Secondary: 15A18.}
\keywords{Out-of-core algorithm, parallel computing, randomized algorithm, singular value decomposition}

\begin{abstract}
Singular value decomposition (SVD) is a standard matrix factorization technique that produces optimal low-rank
approximations of matrices. It has diverse applications, including machine learning, data science and signal
processing. However, many common problems involve very large matrices that cannot fit in the main memory of
commodity computers, making it impractical to use standard SVD algorithms that assume fast random access or large
amounts of space for intermediate calculations. To address this issue, we have implemented an out-of-core (external
memory) randomized SVD solution that is fully scalable and efficiently parallelizable.  This solution factors both
dense and sparse matrices of arbitrarily large size within arbitrarily small memory limits, efficiently using
out-of-core storage as needed.  It uses an innovative technique for partitioning matrices that lends itself to
out-of-core and parallel processing, as well as memory and I/O use planning, automatic load balancing, performance
tuning, and makes possible a number of other practical enhancements to the current state-of-the-art. Furthermore,
by using persistent external storage (generally HDDs or SSDs), users can resume interrupted operations without
having to recalculate previously performed steps, solving a major practical problem in factoring very large
matrices.
\end{abstract}

\maketitle

\section{Introduction}

Singular value decomposition (SVD) is a widely used matrix factorization technique with broad applications. Given a
matrix $A\in\mat{m}{n}$, singular value decomposition consists of a factorization $A=USV^T$, where $U\in\mat{m}{m}$
and $V\in\mat{n}{n}$ are orthogonal, and $S\in\mat{m}{n}$ is a diagonal matrix. The non-zero values of $S$ are called
\emph{singular values}, while the columns of $U$ and $V$ are conventionally called the \emph{left} and
\emph{right singular vectors} respectively.  Given a matrix of rank $n\in\nat$, its best approximation among all
matrices of some fixed lower rank $r<n$ is obtained by SVD.  Given the factorization $A=USV^T$, we can construct a
low-rank approximation $A_r$ of rank $r$ by truncating the diagonal matrix $S$ to rank $r$ and then calculating
$A_r=U S_r V^T$.

In less formal language, SVD is a mechanism for lossy compression.  It calculates a smaller matrix which preserves
as much of the information in the original matrix as possible.  It is often used for data compression, dimensionality
reduction, and signal de-noising as well as a growing list of applications in machine learning and artificial
intelligence.  As SVD applications grow, so does the size of the matrices which we would like to factorize.
Consequently, researchers have developed a number of methods for efficiently performing SVD and for approximating
SVD factorizations.  (For an overview, see Dongarra~\emph{et~al.}~\cite{survey}.)

However, an increasingly important limitation on the use of SVD is more practical: Large, useful matrices do not fit
into the main fast-access memory of reasonably priced computers, and the growth in the sizes of useful data sets
ensures that investing in expensive supercomputers is not a robust solution.  Devising efficient, implementable
out-of-core algorithms for SVD has become a pressing matter with an active research community that has published
various proposals. (See, \emph{inter alia}, Kabir~\emph{et~al.}~\cite{oom}, Haidar~\emph{et~al.}~\cite{oom-ieee},
Rabani~\&~Toledo~\cite{rabani-toledo}, Gates~\emph{et~al.}~\cite{bidiagonal} and with a more narrow focus
on natural language processing, Martin~\emph{et~al.}~\cite{lsa}.)

None of these approaches, however, are suited to SVD approximations. The only previous research we have identified
that addresses out-of-core processing for approximate SVD is Lu, Ino \& Matsushita~\cite{rsvd}, who present an
out-of-core SVD solution based on the \emph{randomized SVD} approximation algorithm developed by Halko, Martinsson
\& Tropp~\cite{hmt}. Their approach, however, uses a graphics card as a fast matrix processor, and treats the memory
on the card as ``core'' and the main memory as ``out-of-core.''  It is not suited to matrices that cannot even fit
into main memory.

Our approach relies on the same \emph{randomized SVD} approximation from Halko, Martinsson \& Tropp as Lu, Ino \&
Matsushita, among other techniques.  However, we present a fully out-of-core solution that assumes matrices are
arbitrarily large, but still fit into conventional persistent storage (like SDDs and HDDs), while main memory is
arbitrarily small.  The \libs supports both full SVD factorization and \emph{randomized SVD} approximation.

The core innovation of the \libs is using \emph{block cyclic data distribution} \cite{osti} to perform all matrix
operations on sub-matrices.  Using this technique, only a small part of any given matrix needs to be in core memory
at any one time, and we can partition arbitrarily large matrices into arbitrarily small sub-matrices and move them
in and out of memory as needed.  Although originally deployed in ScaLAPACK for shared-memory distributed matrix
processing~\cite{Bouge1996}, we find that block cyclic data distribution also serves well as a basis for out-of-core
matrix operations.

We find previous work implementing block-wise out-of-core matrix factorization for LU, QR, and Cholesky methods, but
not SVD.~\cite{oom-ieee}\cite{DAzevedoD00}  All previous out-of-core solutions for SVD that we are aware of have
assumed that at least individual rows and columns can fit into main memory.  Our solution does not have any such
limitations.

This solution implements a number of related enhancements and features designed to increase scalability and improve
on the state-of-the-art, including:

\begin{enumerate}
\item Calculating the amount of memory required for all matrix operations before starting them, and planning memory
usage based on those calculations.
\item Multiple precision support for scalar values - single, double, and in part half-precision floating point
numbers - and row and column indices of either 32 or 64 bits in sparse matrices. We have implemented direct
cross-precision operations for all combinations without preliminary data conversion, except for only partial
half-precision support due to hardware limitations.
\item Compact formats for sparse and dense matrices and support for operations combining the two without conversion in
memory.
\item Matrix data formats that readily support fast multi-threaded data format conversion.  Data conversion at
input and output time is particularly fast, and users can choose to maintain all out-of-core matrix data in
transportable formats if they wish.
\item Automatic balancing of computational resources.
\item Support for persistence and recoverability, so that users can resume interrupted operations with-
out having to recalculate already completed matrix operations.
\end{enumerate}

This solution is also particularly portable and suited to embedded applications, since it has been designed for
to work within fixed, low amounts of memory, and has only a few minor external dependencies, like the standard C++
library, POSIX pthreads, and operating system I/O.

\section{Theoretical background} \label{sec:theory}

Given a matrix $A\in\mat{m}{n}$, singular value decomposition consists of a factorization $A=USV^T$, where
$U\in\mat{m}{m}$ and $V\in\mat{n}{n}$ are orthogonal, and $S\in\mat{m}{n}$ is a diagonal matrix. The non-zero values
of $S$ are called \emph{singular values}, while the columns of $U$ and $V$ are conventionally called the \emph{left}
and \emph{right singular vectors} respectively.

\subsection{Randomized SVD Approximation} \label{sec:rsvd}

The \libs implements the \emph{randomized matrix approximation} technique for SVD described in Halko, Martinsson and
Tropp~\cite{hmt} and Halko~\cite{halko}.  In summary, the algorithm consists of the following steps:

\begin{enumerate} \label{randomized-algorithm}
\item Generate a random Gaussian matrix $O\in\mat{n}{r}$, where $r\ll n$. Each entry is generated independently from
the distribution $N(0,1)$.\label{i:gaussian}
\item Set $Y\in\mat{m}{r}$ such that $Y\as AO$.\label{i:oat}
\item Compute the QR-decomposition of $Y$, i.e., find an orthogonal matrix $Q\in\mat{m}{m}$ and upper-triangular
matrix $R\in\mat{m}{r}$ such that $Y=QR$. Note that the last $m-r$ rows of $R$ will be zero.\label{i:orthogonal}
\item Let $Q_r\in\mat{m}{r}$ be the sub-matrix of $Q$ comprising of the first~$r$ columns of~$Q$, removing the zero
rows of $Q$.\label{i:trunc}
\item Set $B\in\mat{m}{r}$ such that $B\as Q_r^T A.$\label{i:at}
\item \label{i:svdmain} Compute an SVD of $B$, i.e., find $U_B$, $S_B$ and $V_B$ such that $B=U_B S_B V_B^T$.
\item Compute the approximation $U_r\approx U\in\mat{m}{r}$ by $U_r=Q_r U_B$. \label{i:post}
\end{enumerate}

This algorithm requires us to compute an SVD for matrix $B$ with $r$ rows instead of matrix $A$ with $n$ rows, which
is much easier since $r\ll n$, and from which we can recover an approximation through a single multiplication.  The
mathematical basis for this algorithm and its error distribution are discussed at length in Halko, Martinsson and
Tropp~\cite{hmt} and Halko~\cite{halko}.  As a brief qualitative demonstration of this algorithm's approximations,
see Section~\ref{sec:lennatests}.

To compute the SVD of $B$, we have implemented the standard QR-based SVD
algorithm~\cite[Algorithm 8.6.2]{golub-vanloan}, which readily lends itself to parallelization.

\subsection{Power Matrix Technique} \label{sec:powermatrix}

Halko~\cite[1.5.3]{halko} shows empirically that the accuracy of the \emph{randomized SVD} approximation algorithm
depends on the rate of decay of the singular values of the matrix to which it is applied. If this decay is
sufficient, the randomized SVD approximation algorithm will yield adequate results. Otherwise, we apply the
\emph{power iteration} technique described in Halko~\cite[1.6.2]{halko}.

We apply the same \emph{randomized SVD} algorithm to a modified input matrix $A_q\in\mat{m}{n}$ such that:

\begin{equation} \label{eq:powermatrix}
A_q\as \l(AA^T\r)^q A,
\end{equation}

\noindent where $q$ is the \emph{iteration parameter}, generally a small integer value in the range $q=1,\dots,5$.

As demonstrated in Halko~\cite[1.6.2]{halko}, the matrix $A_q$ has the same singular vectors as $A$, and
furthermore, if $\sigma\ge0$ is a singular value of $A_q$, then $\sigma^{\rec{2q+1}}$ is a singular value of $A$.
The choice of iteration parameter $q$ has a substantial impact on the accuracy -- as shown visually in
Section~\ref{sec:experiments} -- but is dependent on the rate of decay of the singular values of $A$.  We lose
accuracy if we set $q$ to low, and we cannot simply set $q$ to a constant high value, because this will make the
singular values decay too quickly, and they will be lost to the approximation.  We can only calculate the optimal
iteration parameter empirically for a given matrix.

One important innovation of the \libs is that it is able to select the best value of the iteration parameter
automatically.  We are able to measure the rate of decay of the singular values of matrices $A_q\in A, A_1, ..., A_n$
with increasing $q$, and identify the value that maximizes accuracy.  We also provide for monitoring the average
$L_2$-norm per matrix item so that we can stop increasing the iteration parameter when the norm falls below a
pre-defined threshold.

Users may also explicitly specify the iteration parameter at run time.

Furthermore, we make significant gains in processing time using the \emph{power iteration} by optimizing the order
of matrix operations.  Note that when performing the first step (\ref{i:gaussian}) of the \emph{randomized SVD}
algorithm, we need to compute:

\begin{equation} \label{eq:powermatrixtimesgauss}
A_q\as \l(AA^T\r)^q A O,
\end{equation}

\noindent where $O\in\mat{n}{r}$ is a random Gaussian matrix. Since $r\ll n$, it is much more efficient to compute
this product from right to left, starting by calculating $AO$ first.  When appropriately optimized, computing power
iterations is nearly an in-place operation and therefore requires little or no additional memory.

\subsection{Block-wise matrix operations} \label{sec:blockprocessing}

Given a pair of matrices $X,Y$ we can calculate their product $A := XY$ by partitioning the two matrices into
quadrants and calculating the products of the individual blocks:

\begin{equation}
\begin{split}
A = &
\begin{pmatrix}
X_{11} & X_{12} \\
X_{21} & X_{22}
\end{pmatrix}
\begin{pmatrix}
Y_{11} & Y_{12} \\
Y_{21} & Y_{22}
\end{pmatrix} \\
= &
\begin{pmatrix}
X_{11}Y_ {11} + X_{12}Y_ {21} & X_{11}Y_ {12} + X_{12}Y_ {22} \\
X_{21}Y_ {11} + X_{22}Y_ {21} & X_{21}Y_ {12} + X_{22}Y_ {22}
\end{pmatrix}
\end{split}
\end{equation}

Using this procedure, we calculate a single matrix product by calculating eight products of matrices one-quarter the
size of $X$ and $Y$.  The total space in core memory needed to obtain these eight products is one-quarter as large as
for the naive matrix multiplication algorithm.  The number of scalar operations needed to calculate the product is
identical to the naive algorithm.

This approach generalizes to more complex and recursive matrix partitions, and draws on work derived from Dongarra
\& Walker.~\cite{osti}  If taken to the most extreme limit, a block does not need to have more than one value in it,
at which point this algorithm becomes identical to ordinary matrix product calculations.  There is no other minimum
block size and therefore, no specific minimum memory needed to calculate the products of arbitrarily large matrices.
We do not even have to keep individual rows or columns entirely in memory.

Just as we can take a block-wise product of two large matrices without keeping more than a small part in core memory
at any one time, we can also factor large matrices in a block fashion, as described in D'Azevedo and
Dongarra~\cite{DAzevedoD00}.  We have implemented all matrix operations in the \libs to operate over sub-matrices,
so that there is no requirement for a whole matrix to ever reside in memory at any one time.

\section{Description of the SVD library} \label{sec:implementation}

The \libs has the following essential features:

\begin{enumerate}
\item Support for:
\begin{enumerate}
\item Approximate and exact SVD of arbitrarily large matrices, much larger than the available main memory.
\item Both sparse and dense matrices, with all matrix operations available for both types without data conversion.
\item Single, double and, with some limitations, half-precision floating-point matrix values.
\item An arbitrary number of processing threads.
\end{enumerate}
\item Efficient switching between in-core and out-of-core processing, so that a single library can apply to
matrices of all sizes.
\item Automatic balancing of computational resources depending on the task's scale.
\item Minimized impact of I/O operations when running in out-of-core mode.
\item Persistence and recoverability, so that users can resume interrupted operations without having to
recalculate already completed matrix operations.
\end{enumerate}

To the best of our knowledge, these features have never been implemented in an SVD library before.

\subsection{Distributed, parallel block-wise processing}

All matrix operations in the \libs are performed over sub-matrices or blocks, as described in
Section~\ref{sec:blockprocessing}.  Blocks are specified as ranges of row and column indices in the source matrix.
Matrices are partitioned into a number of blocks calculated in function of the memory limits provided by the user.

We have implemented sparse matrices so that binary searches provide fast access to values within a range of indices
and can immediately identify block operations that will produce no results because the sparse matrix contains no
values in the appropriate range. For dense matrices, we use pointer arithmetic to directly access any value or range
of values efficiently from its coordinates.  Since accessing the values in matrices is thread-safe, we can easily
distribute and parallelize matrix operations over as many threads as we can efficiently use, without needing to copy
the matrices as long as the threads share memory access to them.

\subsection{Multiple precision arithmetic}

The \libs has full support for both single and double precision values, including support for combinations of the two
without conversion.  We have additionally implemented half-precision values for in-core and out-of-core storage,
but these values are converted to single precision when calculations are performed.  Half-precision value support
enables users to significantly reduce memory and out-of-core space requirements, but do not to save processing time
in this implementation. Current commodity CPUs do not support half precision arithmetic directly at the hardware
level, but CPUs with the \texttt{F16C/CVT16} instruction set do support half-precision values as I/O. Our motivation
for introducing half-precision in the \libs is to enable us to investigate the numerical stability and the quality
of results of randomized SVD with half-precision values, so that we can fully support them in GPUs and future CPUs.
Our half-precision implementation is fully compatible with the \texttt{IEEE 754-2008} standard.

\subsection{Dense and sparse matrices}

All operations in the \libs have either matrices as arguments, or matrices and scalars.  There is no specific class
for vectors, which must be constructed as single row or single column matrices.  However, there is explicit support
for scalar values as a class. They are not treated as $1\times 1$ matrices.

All matrices are either treated as \emph{sparse} or \emph{dense}, with sparse matrices represented in coordinate
(COO) format~\cite{Duff1986}, which is suited especially well for parallel processing.  Row and column indices may be
32 or 64 bit values, to enable compact storage of arbitrarily large sparse matrices.

We have directly implemented efficient matrix multiplication for all four combinations of sparse and dense matrices
(i.e., \emph{dense~$\times$~dense}, \emph{dense~$\times$~sparse}, \emph{sparse~$\times$~dense},
\emph{sparse~$\times$~sparse}) so that no matrix needs to ever be converted from sparse to dense or vice-versa.
Matrices are also marked as in canonical or transposed form, with direct support for all operations in both cases,
so that we do not need to perform memory-consuming transposes during processing and can, in so far as possible,
ensure that matrix row and column reads are linear reads on adjacent values in memory.

By providing individual implementations for each of the four class pairs, we can use the maximally efficient
algorithm for each.  For example, multiplying a very large dense matrix by a very sparse one is proportionate only
to the number of non-zero entries in the sparse matrix.

\subsection{Memory and processing time prediction}

Because we know before computing exactly how many non-zero scalar values there are in a sparse matrix, and can
trivially calculate the number of scalar values in a dense one from its dimensions, we can estimate of the number of
scalar operations and memory required for each individual matrix operation in a product of multiple matrices.  We
can only provide an upper-bounds estimate of the memory and processing required for \emph{sparse~$\times$~sparse}
products, but for all \emph{sparse~$\times$~dense} and \emph{dense~$\times$~dense} combinations, the calculations
are exact.

Using this information, the \libs can plan the order of operations to minimize memory usage and
processing time, i.e., if a matrix product $ABC$ is more efficiently computed as $A(BC)$ than as $(AB)C$, we can
readily identify this fact and proceed appropriately.  We are also able to determine optimal block sizes and predict
when we will need to switch to out-of-core processing without having to lose computing resources to monitoring
memory usage.

We can furthermore calculate with reasonable accuracy the optimal number of threads to use to perform a matrix
operation. We can tune the minimum number of scalar operations a thread must perform to compensate for the fixed
computing resources required to create and run a thread, and since we have already estimated the number of scalar
operations required to perform some matrix calculation, we can easily calculate the proper number of threads.

We are unaware of any other linear algebra suite that provides this kind of predictive memory and thread management.

\subsection{In-core and out-of-core matrices} \label{sub:allocation}

Users control the amount of main memory allocated for a matrix process by setting three parameters:

\begin{enumerate}
\item The amount of memory allocated for each matrix individually.\label{j:1}
\item The amount of memory allocated for new matrices constructed during processing or as output.\label{j:2}
\item A global memory limit for the entire process.\label{j:3}
\end{enumerate}

If the parameter in \ref{j:1} is not set, then the parameter in \ref{j:2} is treated as if it was also the setting
for \ref{j:1}.  Similarly, if the parameter in \ref{j:2} is not specified, then the parameter in \ref{j:3} is used
in its place. If \emph{none} of the parameters are set, the process is treated as entirely \emph{in-core} by
default with the \libs choosing when to use out-of-core processing, unless the user specifies memory parameters at
run-time.

This granular control of memory allocation allows users to, for example, specify that some highly used matrix must
be kept in-core at all times.  If at any stage in processing, any matrix does not fit in the available or specified
memory, it is automatically transferred out-of-core. The matrix I/O handler is small and operates in its own low
priority thread, minimizing its impact on direct matrix processes in memory.

\subsection{Storage structures}

\begin{figure}[h!bt]
\includegraphics[width=100mm]{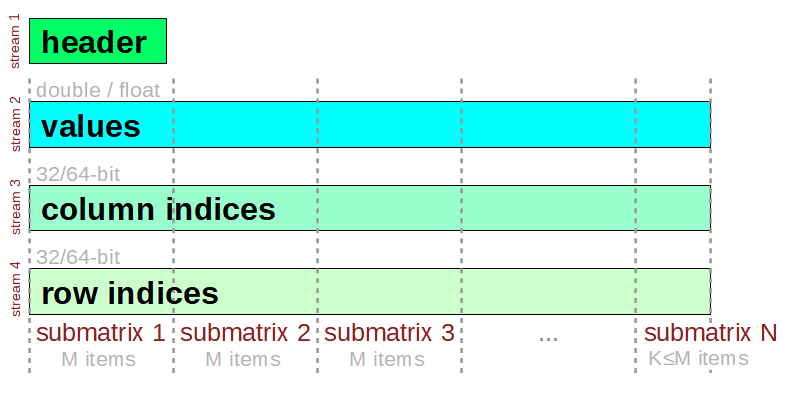}
\caption{Sparse matrix representation}
\label{fig:sparse}
\end{figure}

We store sparse out-of-core matrices in a structure-of-arrays form optimized for fast parallel access. In
particular, row and column indices are stored in a separate array from matrix values, as shown in
Figure~\ref{fig:sparse}. This minimizes mutual data dependencies, which is very important for massive multi-threaded
processing. It has the additional benefit that data and indices can be stored in separate files, bypassing operating
system limitations on file sizes and file reads.

\subsection{Multi-threading}

\libs uses the \emph{POSIX pthreads} execution model to provide efficient process threading.  Users can determine the
number of threads at run-time, or at any stage during processing, including setting the number of threads used for
individual operations or matrices.  To avoid inefficiencies caused by threading overhead, there is an internal
automatic re-balancing mechanism to prevent the generation of large numbers of threads for small tasks.

All I/O is handled via threaded streams. Using structure-of-arrays and COO formats for sparse matrices makes it
possible to read the parts of an out-of-core matrix in multiple parallel threads with little or no I/O latency.
There are also facilities in the \libs for profiling I/O, and some dynamic buffer size management designed to better
balance I/O time and computing time. These enhancements and optimizations enable performance to scale nearly
linearly with the number of threads.

\subsection{Persistence and recoverability}

Among the more innovative features of the \libs is its persistence and ability to resume computing after being
interrupted.  This is a natural extension of out-of-core processing, which keeps intermediate matrices and partial
results in persistent stores.  Every intermediate matrix, including those produced by sub-matrix operations, are
stored with a header indicating their place in the execution plan and relation to other matrices. By recovering
those headers, the library can identify which operations have been completed and only calculate those operations not
yet completed.

This mechanism makes it possible to recover from external or internal failures, as well interrupting a running
operation to re-prioritize tasks or change configurations.

\section{Applications and experiments} \label{sec:experiments}

\subsection{Large Matrix Tests}

To empirically verify the scaling properties of the \lib, we chose two matrices from the \emph{SuiteSparse Matrix
Collection}\footnote{\url{https://sparse.tamu.edu/}} (formerly the \emph{University of Florida Sparse Matrix
Collection)}~\cite{tamu}:

\begin{enumerate}
\item \texttt{Hamrle3}, a $1,447,360\times 1,447,360$ matrix with $5,514,242$ non-zero items
\item \texttt{Circuit5M}, a $5,558,326 \times 5,558,326$ matrix with $59,524,291$ non-zero items
\end{enumerate}

If encoded as a dense matrix, the \texttt{Hamrle3} matrix would occupy more than $15$TB of storage and the
\texttt{Circuit5M} matrix would use roughly $225$TB.  Encoded in sparse matrix form, these matrices are small
enough that we could fit them into the core memory of the computer we used to perform our experiments.  We are
therefore able to show a smooth drop in performance as the memory allocated drops below the minimum for entirely
in-core processing.

\begin{figure}[h!bt]
\begin{subfigure}[b]{0.45\textwidth}
\includegraphics[width=\textwidth]{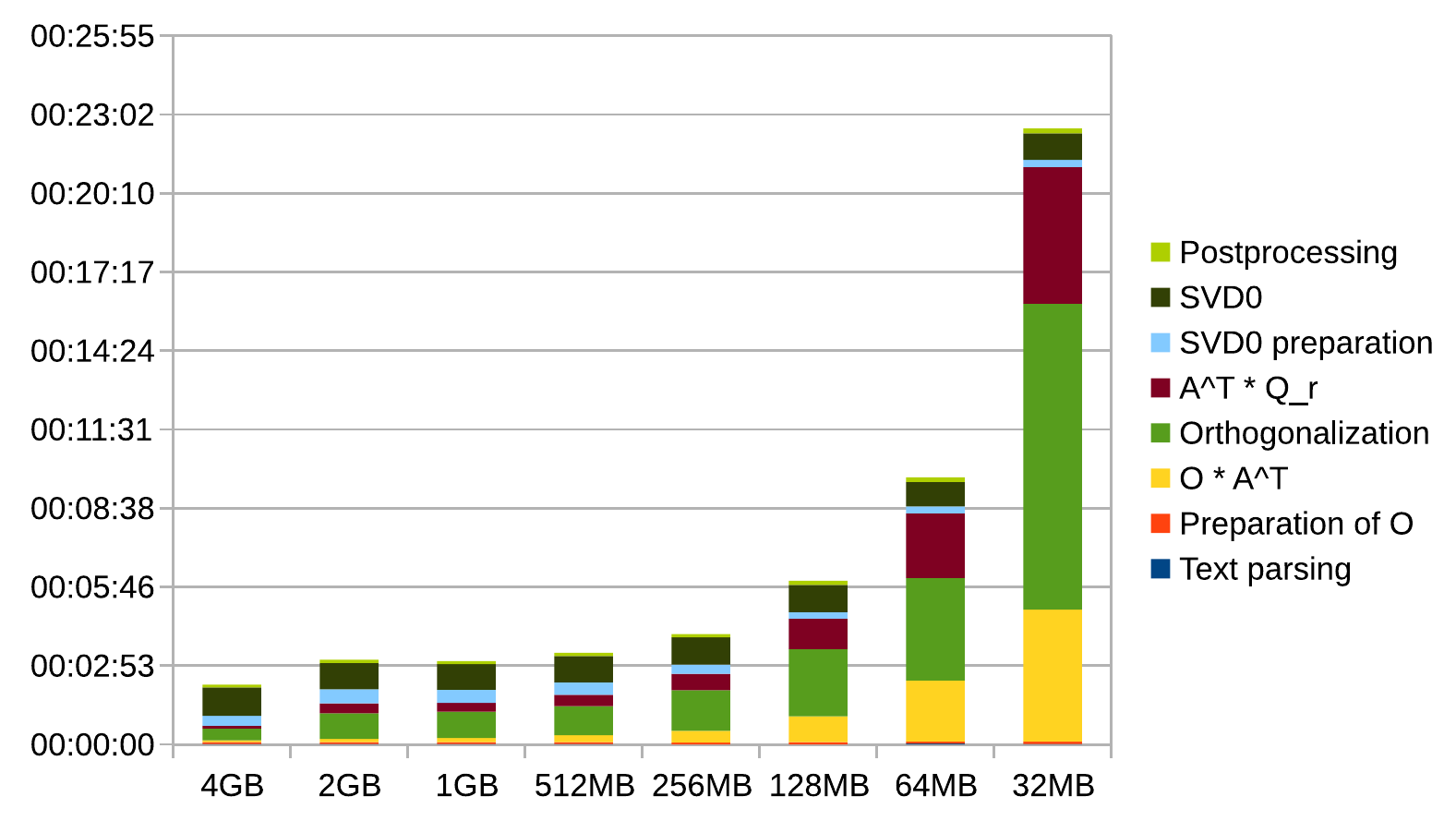}
\caption{Hamrle3}
\label{fig:Hamrle3}
\end{subfigure}
\qquad
~ %
\begin{subfigure}[b]{0.45\textwidth}
\includegraphics[width=\textwidth]{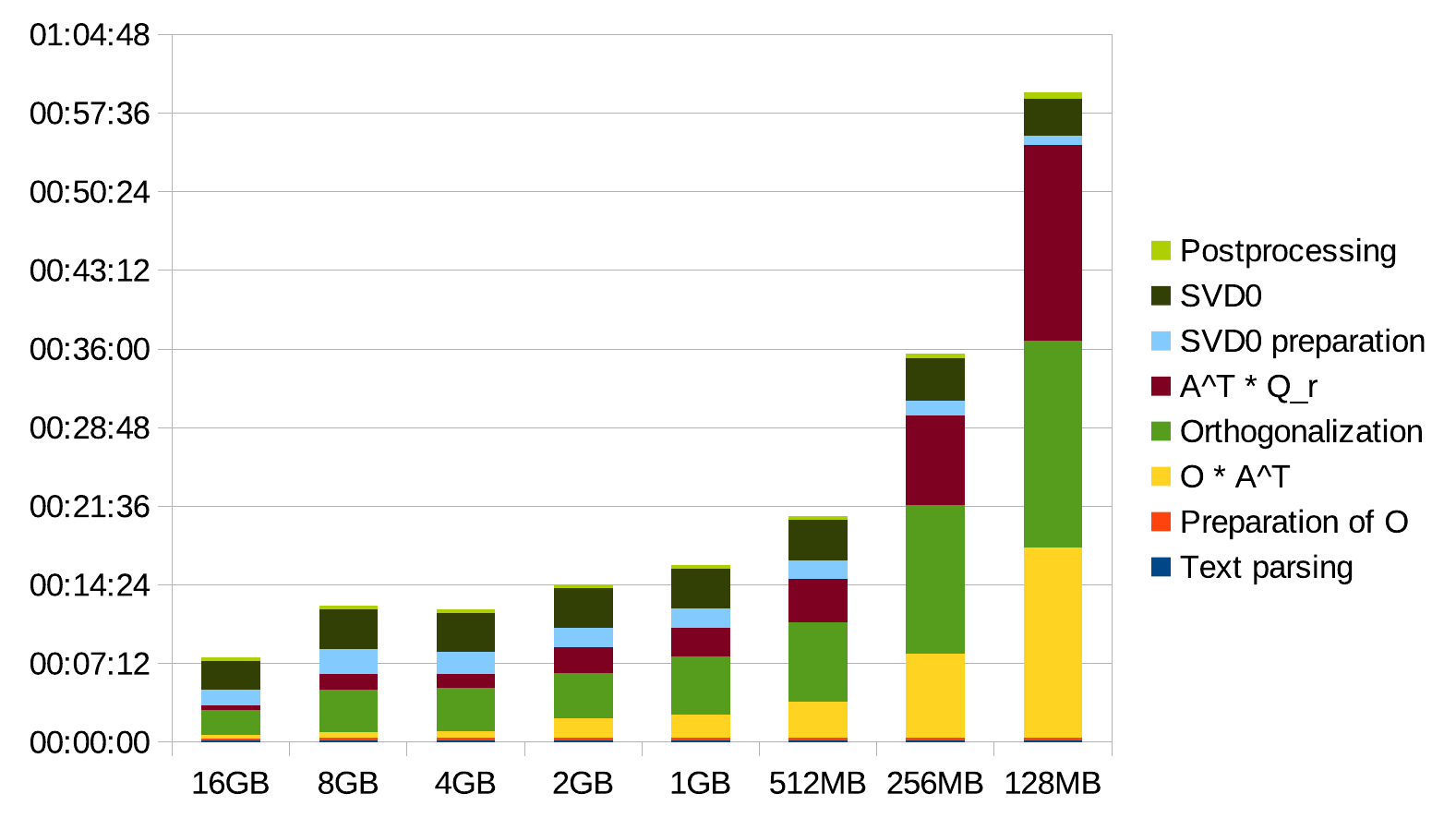}
\caption{Circuit5M}
\label{fig:Circuit5M}
\end{subfigure}
\caption{SVD factorization time as a function of memory limits.  The horizontal axis is the memory limit (per
matrix), and the vertical axis displays the total processing time in \emph{hours}:\emph{minutes}:\emph{seconds}.
Different colors represent different stages of processing.}
\label{time}
\end{figure}

Figure~\ref{time} displays the processing times calculating the approximated SVD factorizations of these two
matrices, as a function of the memory limit we allocated to it.  The experiments were performed on a dedicated
system with dual Intel Xeon E5-2640 v4 @2.40GHz CPUs, $512$GB RAM as core memory, and a $2$TB NVMe-SSD as out-of-core
storage. Memory limits were allocated \emph{per matrix} (see Parameter~\ref{j:1} in Section~\ref{sub:allocation})
and not as a global memory limit.

In the slowest case - factoring the \texttt{Circuit5M} matrix in $128$MB of core memory - the test took roughly one
hour.

Times in Figure~\ref{time} are divided into different parts of the factorization task:

\begin{enumerate}
\item \texttt{Text Parsing}: The matrices are stored in \emph{Matrix Market Exchange}
format\footnote{\url{https://math.nist.gov/MatrixMarket/formats.html}}~\cite{MatrixMarket}, which must first be
parsed and stored in the \libs sparse matrix representation.  This did not consume a large share of total processing
time.
\item \texttt{Preparation of $O$}: Generating the Gaussian (random) matrix $O$ (see Step~\ref{i:gaussian} in
Section~\ref{randomized-algorithm}). This includes allocating in- and out-of-core space for it, and writing to the
out-of-core store.
\item The next three steps -- \texttt{O*A\textasciicircum T} ($OA^T$), \texttt{Orthogonalization}, and
\texttt{A\textasciicircum T*Q\textunderscore r} ($A^T  Q_r$) -- encompass respectively Steps~\ref{i:oat}, \ref{i:orthogonal}
and \ref{i:trunc}$/$\ref{i:at} in Section~\ref{randomized-algorithm}.
\item \texttt{SVD0 Preparation}: Before performing SVD calculation in Step~\ref{i:svdmain}, we must transpose the
matrix $B^T\as A^T  Q_r$ to obtain $B$, and then compute $B^T B$. Included in the time allotted to this part of the
process are some other minor memory operations in preparation for the full SVD calculation.
\item \texttt{SVD0}: We then perform the full SVD calculation. (Step~\ref{i:svdmain} in Section~\ref{randomized-algorithm}.)
\item \texttt{Postprocessing}: Finally, we compute approximate $U,S,V^T$. (Step~\ref{i:post})
\end{enumerate}

In Figure~\ref{time}, the runs with the most memory allocated to the \libs are, in both cases, instances of entirely
in-core processing.  The \texttt{Hamrle3} matrix can be processed with an allocation of $4$GB and the
\texttt{Circuit5M} matrix in $16$ GB. The \libs recognizes that with this large an allocation it can perform the
entire factorization in memory, and never has to unload data from core memory and reload it from external stores.

In all other cases, out-of-core processing is used for at least some portion of the factorization task.  However,
the second bar in both charts represents cases where \emph{in fact} the entire matrix resides in memory, but is
divided unequally between two blocks in order to use out-of-core processing if needed.  Block-wise processing in
this case increases run-times significantly in order to switch between the blocks and reassemble the results of
operations.  Because the blocks are not equal in size, this is relatively inefficient.

The third bar in both charts shows slightly \emph{less} run-time even though less memory has been allocated.  In this
case, matrices are partitioned into more parts with a better balance in sizes.  Although in this case the input
matrices remain in memory, processing remains slower than in the case where ample memory is allocated and the
library does not prepare for any out-of-core processing.

As the number of matrix parts increases due to a decreasing memory limit, the time spent performing matrix
multiplications grows as a share of total processing time, reflecting the degree to which time must be spent
switching between blocks and, if the process is out-of-core, performing I/O.  This is the principal source of the
trade-off between memory allocation and processing time. Unsurprisingly, the most important bottleneck to out-of-core
processing is I/O.  With the use of fast SSDs, this bottleneck is less important than in the recent past.  Moreover,
we have made considerable efforts to streamline and optimize I/O in \lib, which has clearly had a positive impact.

Nonetheless, we clearly show that in our implementation, this trade-off is sub-linear. In both charts, the memory
allocated to the process reported in the last bar is 128 times smaller than in the first bar, but processing time
has increased by a factor of less than 10.

\subsection{Qualitative performance of randomized SVD} \label{sec:lennatests}

Although our intended use for the \libs is in artificial intelligence and natural language processing, we present
here a test application in image compression to provide a qualitative assessment of \emph{randomized SVD} as an
approximation method and of the  \emph{power iteration} technique to improve the quality of the result.

We use the standard test image known as ``\emph{Lenna}''\footnote{For the story of this well-known image, see
\url{http://www.lenna.org}.}, a $512\times512$ pixel image cropped from a scan of an image published in
\cite{lenna}.  We treat this image as an 8-bit greyscale bitmap, structured as a matrix $A\in\mat{512}{512}$, with
pixel brightness stored as a single, double, or half precision value, depending on the test setup.

To compress the image, we perform \emph{randomized SVD} approximate factorization on the image matrix $A=USV^T$,
where $U,V\in\mat{512}{50}$ and $S\in\mat{50}{50}$ is a diagonal matrix.  Multiplying the matrices out produces a
$512\times512$ matrix that we can convert back into an 8-bit greyscale bitmap and display as a reduced fidelity
version of the original image.

These three matrices together contain $51,250$ values, with the total number of bytes, and thus total compression,
dependent on the precision. The choice of rank 50 is somewhat arbitrary and chosen because at higher rank there
was little or no visible difference between the compressed and uncompressed images in a printed medium.

For the purposes of these demonstrations, we performed no pre- or post-processing, re-scaling, or normalization of
the source image or compressed form.

\subsubsection{Power iteration}

\begin{figure}[h!bt]
\begin{subfigure}[b]{0.3\textwidth}
\includegraphics[width=\textwidth]{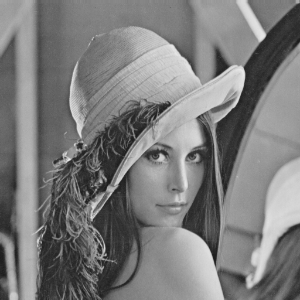}
\caption{Original image. \\ ~}
\label{fig:1}
\end{subfigure}
\qquad
~ %
\begin{subfigure}[b]{0.3\textwidth}
\includegraphics[width=\textwidth]{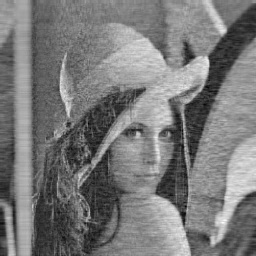}
\caption{Rank~$50$ compression,\\ no power iterations.}
\label{fig:2}
\end{subfigure}
\qquad
~ %
\begin{subfigure}[b]{0.3\textwidth}
\includegraphics[width=\textwidth]{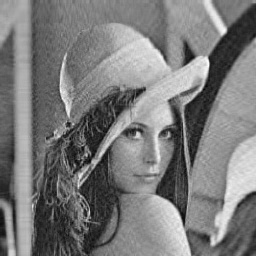}
\caption{Rank~$50$ compression,\\ with power iterations.}
\label{fig:3}
\end{subfigure}
\caption{Image reconstruction via low rank SVD with and without power iterations.}
\label{lenna}
\end{figure}

Figure~\ref{lenna} is a demonstration of the effectiveness and stability of the \emph{randomized SVD} approximation
algorithm (Section~\ref{sec:rsvd}) and the \emph{power iteration} technique (Section~\ref{sec:powermatrix}) as
applied to this image compression task.  The quality of the compressed image is notably better using
\emph{power iteration} with an automatically calculated optimal iteration parameter (Figure~\ref{fig:3}) than using
no \emph{power iteration} at all (Figure~\ref{fig:2}).  This result offers support for the empirical observations
about this technique in Halko~\cite{halko}.

\subsubsection{Precision and memory usage}

The source image has dimensions of $512\times512$ pixels, with one byte of greyscale information per pixel.  This
amounts to $262,144$ bytes of memory to store this as a bitmap. The total number of scalar values in the three
matrices used for the rank 50 compressed version is $51,250$. The number of bytes required to store these matrices
depends on the precision of their values - from $102,500$ bytes for half precision, to $410,000$ bytes at double
precision.  Given that the 8-bit images are already massively less precise than even half precision values, there is
no need to be concerned about this as an image compression question.  However, the choice of precision does have an
impact on the output of the \emph{randomized SVD} approximation algorithm as well as on memory usage during
processing.

\begin{figure}
\begin{subfigure}[b]{0.3\textwidth}
\includegraphics[width=\textwidth]{./pics/50-lenna-double.jpg}
\caption{Double precision.}
\label{fig:double}
\end{subfigure}
\qquad
~ %
\begin{subfigure}[b]{0.3\textwidth}
\includegraphics[width=\textwidth]{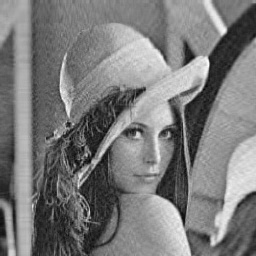}
\caption{Single precision.}
\label{fig:single}
\end{subfigure}
\qquad
~ %
\begin{subfigure}[b]{0.3\textwidth}
\includegraphics[width=\textwidth]{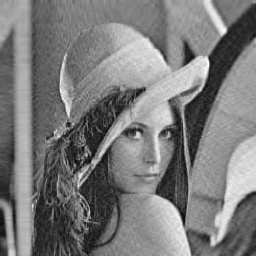}
\caption{Half precision.}
\label{fig:half}
\end{subfigure}
\caption{Rank $50$ image reconstruction with different precisions.}
\label{lenna-precision}
\end{figure}

Figure~\ref{lenna-precision} displays the effect of using different precisions to perform the same rank $50$
approximation on the \emph{Lenna} image.  You can see that there is little visible quality difference between the
results using different precisions.  We see no evidence that reduced precision undermines the \emph{randomized SVD}
approach, and since each reduction in precision from double to single to half reduces the memory required for
processing by 50\%, the gains from making fuller use of low precisions appear to be worth consideration.  Stability
with half-precision values indicates that there are large gains in processing speed to be made in providing GPU
support for this algorithm, at very little cost in result quality.  Robustness under variable precision further
suggests that \emph{randomized SVD} is also robust against noisy inputs.

All image compression experiments in this section were performed with the memory allocation restricted to $1$KB per
matrix -- less than the memory required for a single row or column of the source image -- in order to demonstrate
the implementation's ability to function under very extreme memory restrictions.

\section{Conclusions} \label{sec:conclusions}

SVD is time-honored and well-established data processing tool. Even though there are already many numerical SVD
algorithms, new challenges are arising in connection with large scale machine learning and increasingly big data.
Motivated by various applications in artificial intelligence and data science, we have implemented a true
out-of-core scalable SVD solution -- the \libs -- to handle very large datasets.

Furthermore, this library can be equally well applied to perform SVD factorization for relatively small matrices
on machines with very limited core memory. It efficiently supports a number of practical resource-saving techniques,
like variable precision scalar numbers, both sparse and dense matrices, and different levels of approximation in the
result.

\libs implements a sophisticated memory prediction mechanism that can decompose matrix operations into arbitrarily small
tasks, preventing memory exhaustion and allowing permissive task scheduling and prioritization. Among other gains, this
means that \libs can run efficiently in a \emph{docker} container\footnote{\url{https://www.docker.com/}} or some other
virtualized computing environment.  It is also very fail-safe, since it preserves partial results and processing can be
interrupted and resumed at little cost. These are, to the best of our knowledge, new features that have not previously
been implemented in an SVD library.

The automatic selection of the power iterator $q$ (Section~\ref{sec:powermatrix}) is an important advance in numerical
methods for SVD, whose effects are visible in Figure~\ref{lenna} when used as an image compressor.

The \libs is not a new algorithm for performing SVD, and does not change the fundamental computing requirements for
matrix operations, but it does enable users to make efficient trade-offs between computing time, really available
memory and storage, and final accuracy.  This makes it possible to perform even very large matrix decomposition tasks
on systems with very few resources, which has not previously been possible.

To extend the immediate usability of \libs, we have provided a Python NumPy-like interface as well as a Java interface,
in addition to the native C++ API.  The architecture of \libs makes it possible to extend it to support computational
accelerators such as GPUs and FPGAs. We have not provided such support in the current version because the main
bottleneck in out-of-core computing is the I/O system.  We plan to include support for hardware accelerators in
future versions.

The \libs relies on the \texttt{ExB BLAS MLCore Library} \cite{exbblas}, which is used in other higher-level machine
learning task such as clustering, decision tree learning, and approximate and exact \emph{kNN} indexing. Because of its
central role in multiple machine learning tasks, we are continuing work on its further optimization in conjunction with
the \libs.

\section*{Acknowledgments}

We would like to thank our colleagues at \texttt{ExB} for their support during this work, and particularly Scott
Martens for his helpful comments and assistance with writing this paper.  This work was financially supported by
\texttt{ExB Labs GmbH}\footnote{\url{https://www.exb.de/}}.

\bibliographystyle{siam}

\end{document}